\documentclass{aastex} \usepackage[numberedappendix]{emulateapj5}

\shorttitle{TRGB and RC Distance to M33} \shortauthors{Kim et al.}

\begin{document}

\title{
DETERMINATION OF THE DISTANCE TO M33 BASED ON THE TIP OF THE RED 
GIANT BRANCH
AND THE RED CLUMP$^1$}
\altaffiltext{1}{Based on observations with the NASA/ESA Hubble Space
Telescope obtained at the Space Telescope Science Institute, which is
operated by the Association of Universities for Research in Astronomy,
Incorporated, under NASA contract NAS5-26555.}

\author{Minsun Kim, Eunhyeuk Kim, Myung Gyoon Lee}
\affil{Astronomy Program, SEES, Seoul National University, Seoul, 
151-742, Korea}
\email{mskim@astro.snu.ac.kr,ekim@astro.snu.ac.kr,mglee@astrog.snu.ac.kr}

\author{Ata Sarajedini}
\affil{Department of Astronomy, University of Florida, P.O.Box 112055,
Gainsville, FL 32611, USA} 
\email{ata@astro.ufl.edu}

\author{Doug Geisler}
\affil{Departamento de F{\'\i}sica, Grupo de Astronom{\'\i}a, 
Universidad de Concepci\'on,
	Casilla 160-C, Concepci\'on, Chile}
\email{doug@kukita.cfm.udec.cl}

\begin{abstract}
We have determined the distance to M33 using the tip of the red giant 
branch (TRGB) and the red clump (RC), from the $VI$ photometry of stars in ten 
regions of M33
based on $HST/WFPC2$ images.
The regions used in this study are located 
at $R=2.6 - 17.8$ arcmin from the center of M33.
The distance modulus to M33 obtained in this study,
for an adopted foreground reddening of $E(B-V)=0.04$, is
$(m-M)_{0,TRGB}=
24.81\pm0.04$(random)$^{+0.15}_{-0.11}$(systematic)
from the TRGB, and $(m-M)_{0,RC} = 
24.80\pm0.04$(random)$\pm0.05$(systematic)
from the RC, showing an excellent agreement between the two
(corresponding to a distance of $916\pm17$(random) kpc and
$912\pm17$(random) kpc, respectively).
These results are 
$\approx 0.3$ mag larger than the Cepheid distances
based on the same $HST/WFPC2$ data and ground-based data.
This difference is considered partially due to the uncertainty
in the estimates of the total reddening for Cepheids in M33.

\end{abstract}


\section{INTRODUCTION}

The accurate determination of the distances to Local Group galaxies
is critical in the study of the extragalactic distance scale. In 
particular
two spiral galaxies in the Local Group, M31 and M33, are primary 
calibrators
for several secondary distance indicators including  the Tully-Fisher 
relation.
While the distance to M31 has been measured extensively, relatively 
less attention has
been paid to the distance determination of M33 \citep{van00}.

Recently the tip of the red giant branch (TRGB), a Population II 
distance indicator, has frequently been used for the determination 
of distances to resolved galaxies in the
Local Group; in addition, another Population II distance indicator, 
the red clump (RC), has come to be used quite often as well.
These two methods have an advantage over the classical primary 
distance
indicators such as Cepheids and RR Lyraes in that
the distances to galaxies can be determined reliably from single 
epoch observations.
The TRGB has been known to be an excellent standard candle for 
resolved galaxies, because the $I$-band magnitude of the TRGB for 
stars older than a few Gyrs
with metallicity $-2.1\le[Fe/H]\le-0.7$ is essentially independent of 
age and metallicity
\citep{lee93, lee96, sal98, fer00, cio00}.

On the other hand, the RC was not widely recognized as a potentially 
good standard candle
until \citet{pac98} recently suggested its use.
Since then the reliability of the RC as a standard candle has been 
controversial
(see \citet{sar99}; \citet{gir01}; \citet{pop00} and the references 
therein).
\citet{pac98} showed
that the mean $I$-band magnitude $M_{I}^{RC}$ of red clump stars is 
independent of
their color in the range $0.8\le(V-I)_{0}\le 1.4$ and has a small
dispersion($\approx 0.2$ mag), suggesting that the RC can be
a good standard candle as originally pointed out by \citet{can70}.
However,
the distance to the Large Magellanic Cloud obtained using the RC
by \citet{sta98} and \citet{uda98a} was much shorter than that based 
on other methods,
which became a starting point for debate focused on the accuracy of the RC 
method.
\citet{uda98b} claimed that the $M_{I}^{RC}$ has a weak dependence on 
metallicity
and no dependence on age for an intermediate age population (2--10 
Gyrs) of stars.
On the contrary,
\citet{sar99} analyzed several galactic open clusters with 
intermediate-ages and compared the properties of their RCs with the
predictions of  stellar evolution models; this comparison indicates
that the $M_{I}^{RC}$ becomes significantly fainter as the 
cluster gets older.
In addition, a number of studies have argued that stellar evolutionary 
theory predicts a significant dependence of $M_{I}^{RC}$ on the 
combination of the age and metallicity of the stellar population
 \citep{col98, gir98, gir00, gir01}.
\citet{cas00} pointed out that the input physics and the dependence 
of various evolutionary codes should be considered to clarify this 
discrepancy.
Observationally it is also important to attack this problem
by investigating the effect of age and metallicity on the RC in other 
nearby galaxies
where the properties of the RC can be studied in detail.

In this paper we present an analysis designed to determine the 
distance to M33 using the TRGB and the RC based on photometry of 
stars in ten M33 fields obtained from $HST/WFPC2$ images.
M33 is an ideal target to apply both the TRGB and RC methods of 
distance determination using $HST/WFPC2$ data.

This paper is composed as follows.
In \S 2 we present the data and reduction technique.
\S 3 displays the color-magnitude diagrams of the measured stars, 
and estimates the distance to M33 using the TRGB and RC methods.
Primary results are discussed in \S 4 and are summarized in \S 5.

\section{DATA AND REDUCTION}

We have analyzed $HST/WFPC2$ data for ten fields in M33
obtained for \citet{sar98}'s cycle 5 program (GO-5914).
Each field was observed for four orbits, yielding
a total exposure time of 4800 seconds for $F555W (V)$ and 5200 
seconds for $F814W (I)$.
These data were obtained originally for the study of globular 
clusters in M33, and
a globular cluster is centered in each PC chip. The data from the PC 
chip were
presented by \citet{sar98} and \citet{sar00}.
In this study we use all field stars in the WF2, WF3, and WF4 chips 
as well as in the
PC chip.
Hereafter we refer to each observed region using  the globular 
cluster's designation. 
Figure 1 illustrates the location of the regions in M33 used in this 
study.
Considering the number of fields and the deep exposures, these data 
are
ideal for studying the field stars as well as the globular clusters 
in M33.

Table 1 lists the positions, galactocentric distances, deprojected 
radial distances,
and the reddening values of
all the regions used in this study.   
The position of the region in Table 1 is the center of the WFPC2.
The galactocentric distance is the distance from
the center of M33 (RA(2000) = 01$^h$33$^m$51$^s$.02,
Dec(2000) =+30$^\circ$39$^\prime$36$^{\prime\prime}$.7)
\citep{cot99}.
All the regions were assumed to be in the plane of M33's disk
and were deprojected to estimate the actual radial distance.
An inclination of $56^\circ$ and a position angle of
$23^\circ$ for M33 were used for deprojection of the positions 
\citep{reg94}.
For foreground reddening correction, the COBE/IRAS extinction maps
of \citet{sch98} are used. The reddening values of all the regions
are as low as $E(V-I)=0.06$ ($E(B-V)=0.04$).
The extinction laws for $R_V=$3.3, $A_{I}=1.95E(B-V)$ and 
$E(V-I)=1.35E(B-V)$
\citep{car89}, are adopted in this study.

The photometry of the stars in the CCD images
has been obtained using the {\it multiphot} routine of the HSTphot package
\citep{dol00a}. 
The HSTphot package was designed for photometry of $HST/WFPC2$ data and 
employs
a library of Tiny Tim  point-spread-functions (PSFs) for PSF fitting
to account for variations in the PSF due to location on the
chip and the centering within a pixel.
After PSF-fitting, corrections
are also made for geometric distortion, CTE effect,
and the 34$^{th}$ row effect \citep{dol00b}. 
The {\it multiphot} routine gives the magnitudes
transformed to the standard system as well as instrumental magnitudes.
 The HSTphot photometry used zero points
from \citet{dol00b} which provides corrections to the \citet{hol95} values.

\section{RESULTS}

\subsection{Color-Magnitude Diagrams}

The number of the measured stars in each region 
is many tens of thousands (from $\sim 60000$ to $\sim80000$ stars)
which are too many to plot in a color-magnitude diagram (CMD).
Therefore, as an example, Fig. 2 shows the
color-magnitude diagram (CMD) for one field.
In the case of the PC chip, at the center of which a globular 
cluster is located,
the stars at $r<2.8$ arcsec from the center of the cluster are 
considered to be members while those at $r>4.6$ arcsec are 
considered to be field stars.
Figure 2 shows a CMD for the measured stars in the 
C20-region, which happens to be our most distant field
from the center of M33.
Several features are seen in Figure 2.
(a) There is  a broad red giant branch (RGB),
the tip of which is seen at $I \approx 21.0$ mag.
The mean color of the RGB of these field stars is redder
than that of the globular cluster C20 in the same region (represented 
by the solid line). 
The locus of C20 was derived from the median color of the stars 
at  $r<2.8$ arcsec from the center of C20.
(b) A red clump is distinctively seen at $I \approx 24.5$ mag and 
$(V-I) \approx 1.0$;
(c) Asymptotic giant branch (AGB) stars are also seen along and above the 
RGB; and
(d) There is a blue plume  at $(V-I)\approx 0.0$ extending up to $I\approx 20$ mag,
which consists of massive main sequence stars and evolved supergiants.
These features are also seen in the CMDs of the other regions.

In Figure 3, the CMDs of all the regions
are shown in number density contour maps (Hess Diagrams).
The density contour maps are constructed on
$100 \times 100$ grids in the CMD domain (the size of each grid is
$\Delta(V-I) \times \Delta I = 0.03 \times 0.1$),
and are smoothed using Gaussian filters of $2$ grid width.
Basic features seen in Figure 3 are similar to those in Figure 2.
Number density contour maps are useful for revealing the areas
with the highest stellar density.
All CMDs in Figure 3 show a strong peak at the position of the red 
clump (marked by the crosses).
The photometry of the stars in the R14- and R12-regions is significantly 
affected by crowding,
because they are located close to the center of M33.
Therefore only the stars in the PC chip 
(which has higher spatial resolution  than the WF chips) 
are used to measure the magnitude of the red clump in these regions
in the following.

\subsection{Estimation of the Distance to M33}

\subsubsection{Tip of the Red Giant Branch}

We have determined the distance to M33 using the $I$-band magnitude 
of the TRGB,
following the description given in \citet{lee93}.
Figure 4 displays the $I$-band luminosity functions of red stars 
including
the RGB and AGB stars.
In Figure 4 there is a sudden increase at $I_{TRGB} \approx 20.9$ in 
all the
regions as marked by the arrow, which corresponds to the TRGB.
We have measured the $I$-band magnitude of the TRGB using this 
feature in the luminosity function by eye detection
(supplemented by using the edge-detection filters \citep{lee93}).
The values of the $I_{TRGB}$ thus derived for all the regions are 
listed in Table 2.

The distance modulus is given by

\begin{equation}
(m-M)_{I} = I_{0, TRGB} + BC_{I} - M_{bol, TRGB}
\end{equation}
where $I_{0, TRGB}$ is the dereddened $I$-band magnitude of the TRGB.
$BC_{I}$ is the bolometric correction to the I magnitude which 
depends on color as follows:
\begin{equation}
BC_{I} = 0.881-0.243(V-I)_{0, TRGB}
\end{equation}

\noindent where $(V-I)_{0, TRGB}$ is the dereddened color of the TRGB.

The bolometric magnitude of the TRGB, $M_{bol,TRGB}$,
is given as a function of metallicity [Fe/H] by:

\begin{equation}
M_{bol,TRGB} = -0.19[Fe/H] -3.81.
\end{equation}

Metallicity can be estimated from the $(V-I)$ color
at the absolute $I$-band magnitude of $M_I = -3.5$ given by 
\citet{lee93}
(see also \citet{sav00}) as follows:

\begin{equation}
[Fe/H] = -12.64+12.6(V-I)_{0, -3.5} -3.3(V-I)^2_{0, -3.5}.
\end{equation}

We have employed an iterative procedure in which an initial guess at 
the distance is used to estimate the metallicity which is in turn 
used to refine the distance until the solution converges, which occurs 
after only a few iterations.
It is important to note that
the regions used for this study are located in various environments 
including young to old stellar populations; thus,
the broad RGBs seen in the CMDs are actually a mixture of 
intermediate-age to old
populations, as well as a range of metallicities.
If we simply use the mean color [$(V-I)_{0, -3.5}$]
of the entire apparent RGB in this case,
the resulting metallicity will be an underestimate, because
there are younger populations with bluer color on the blue side of 
the RGB. For this reason we tried to use the median value of the color 
of the stars along the RGB to reduce the effect of intermediate-age 
populations.
As a check of our method, we have also derived the mean metallicity 
using the slope of the RGB as calibrated by \citet{sar00}, obtaining 
very similar results to those from the median color of the RGB stars.

The mean metallicities resulting from this procedure are listed in Table 2.
The mean metallicity ranges from [Fe/H] $\approx$ --0.6 
to --0.9 dex.
Figure 5 displays the mean metallicity versus the deprojected 
radial distance of the regions (filled circles). 
In Figure 5 there is clearly a negative radial gradient of the 
metallicity.
The mean metallicity data are fit by
[Fe/H] $= -0.05[\pm0.01] R_{dp} - 0.55[\pm0.02]$
([Fe/H] $= -0.04[\pm0.02] R_{dp} - 0.51[\pm0.06]$, using the
metallicity obtained with the RGB slope method)
for all the data,
where $R_{dp}$ is given in terms of kpc 
($1^\prime=0.27$ kpc is assumed).
If we exclude the two innermost regions where the crowding is severe, 
we obtain a fit,
[Fe/H] $= -0.07[\pm0.01] R_{dp} - 0.48[\pm0.04]$
([Fe/H] $= -0.08[\pm0.03] R_{dp} - 0.34[\pm0.10]$  using the
metallicity obtained with the RGB slope method), 
similar to values found in our Galaxy's disk using open
clusters and field giants ($d[Fe/H]/dR = - 0.050 \pm 0.008$ kpc$^{-1}$)
\citep{jan79}.

In Figure 5 the metallicity of the red giants is compared with that 
of HII regions in M33.
The metallicity of the HII regions was converted from
[O/H] values given in the literature 
\citep{kwi81, mcc85, vil88, zar94} using the following relation
taken from \citet{kin00}: $[O/Fe]=-0.184[Fe/H]+0.019$.
The deprojected radius for the HII regions was calculated as above.
The metallicity of the HII regions is fit
by [Fe/H] $= -0.12[\pm0.02] R_{dp} + 0.33[\pm0.07]$, which
is somewhat steeper than that for the field red giants. 
The relation for the HII regions shows a trend that is similar to
that of the field red giants, but with a 
larger scatter.
It is natural that the mean metallicity of the field red giants is 
lower than that of the HII regions,
because the red giants are much older than the HII regions.

Then we derive the distance modulus of each region using the 
information given above.
Table 2 lists the parameters related to the TRGB method;
the observed $I$-band magnitude of the TRGB ($I_{TRGB}$),
the extinction corrected $I$-band magnitude of the TRGB ($I_{0,TRGB}$),
the mean color of the TRGB [$(V-I)_{0,TRGB}$],
the mean color measured at $M_I = -3.5$ [$(V-I)_{0, -3.5}$],
the mean metallicity ([Fe/H]) of the RGB,
the absolute magnitude of the TRGB ($M_{I, TRGB}$),
and the distance modulus [$(m-M)_{0}$].

Figure 6 displays $I_{TRGB}$ versus [Fe/H] for the ten regions in M33.
The value of  $I_{TRGB}$ varies little, with no obvious net metallicity
dependence.
The mean value of $I_{TRGB}$ for the ten regions is $I_{TRGB} = 
20.88\pm0.04$,
showing a remarkably small dispersion.

The average value of the distance moduli for all of the fields is  
calculated to be 
$(m-M)_{0,TRGB}=24.81\pm0.04$(random)$^{+0.15}_{-0.11}$(systematic).
The errors for the distance modulus are based on the error budget
listed in Table 3.

The calibration of the TRGB is based on Galactic globular 
clusters with
[Fe/H] = --2.1 to --0.7 dex, yet the derived mean metallicities of 
the four inner regions in our
sample ([Fe/H] = --0.61 to --0.68 dex) are slightly larger
than the upper boundary of the calibration range. 
If we use only the six other regions, excluding these four inner ones,
we obtain an average distance modulus of 
$(m-M)_{0,TRGB}=24.83\pm0.06$(random)$^{+0.15}_{-0.11}$(systematic).
If the theoretical calibration given by \citet{sal98} is adopted
($M_{I,TRGB} = -3.953 + 0.437[M/H] + 0.147[M/H]^2$ and
$[M/H]=-39.270+64.687[(V-I)_{0,-3.5}] -36.351[(V-I)_{0,-3.5}]^2 
+6.838 [(V-I)_{0,-3.5}]^3 $),
the average distance modulus
will be $(m-M)_{0,TRGB}=24.99\pm0.04$(statistical), which is 0.2 mag 
fainter than
that derived using the empirical calibration of \citet{lee93}. 
However, it has been found that the distance obtained with the 
theoretical
calibration is not consistent with other results as shown by
\citet{dol01} in IC 1613.
So we prefer to use the empirical calibration rather
than the theoretical one in this study.
Finally  we adopt 
$(m-M)_{0,TRGB}=24.81\pm0.04$(random)$^{+0.15}_{-0.11}$(systematic)
as the TRGB distance modulus to M33.

\subsubsection{Red Clump Stars}

We have determined the distance to M33 using the red clump as well.
Red clumps are clearly found in the CMDs of all the regions, as
seen in Figure 3.
It is important to note that implicit in the subsequent analysis is the 
assumption that the metallicity determined from the RGB stars can be 
applied to the red clump stars as well. This is based on two assertions. 
First, based on the absence of a vertical structure of stars blueward 
of the RC \citep{gir01}, we claim that the age of the RC stars is 
likely to be older than $\sim$1 Gyr. Second, the chemical enrichment in 
the disk of M33 is assumed to have 
been small in the period between 10 Gyr and a few Gyr ago. 

To check the variation of the mean magnitude of the red clump on the 
color,
we have selected red clump stars with colors and magnitudes in the range 
$0.6 < (V-I) < 1.6$ and $23.5 < I < 25.5$.
Figure 7 displays the mean magnitude of these stars ($I_{RC}$) versus
their $(V-I)$ colors.
The horizontal bars on the filled circles represent the size of
the color bin. Two thousand stars are included in each color bin.

Figure 7 shows that the variation of the mean magnitudes for the
color range of $0.8 \le (V-I) \le 1.2$ in a given region
(the same range as used for M31 by \citet{sta98a}) is smaller than 
0.1 mag in all of the regions; Fig. 7 also indicates that
the variation of the mean magnitudes among the regions is 
remarkably small (less than 0.1 mag).
It is rather surprising that the variation of the mean magnitudes of the RC 
is smaller than 0.1 mag, considering that the regions used are 
located at a large range of
distances from the center of M33 with diverse star 
formation histories and a varying metallicity. However, we note that 
the metal abundances vary over a range of less than 0.3 dex. Such a 
small abundance range has a correspondingly small effect on the I-band 
RC luminosity (less than 0.04 mag)\citep{gir01,sar99}. The RC is also 
sensitive to age, which further suggests that the dominant
stellar populations in our M33 disk fields are all probably similar in 
age. In uniform populations such as this, 
the red clump can potentially be a good standard candle.

We derive then the $I$-band luminosity functions of these red clump 
stars with $0.8 \le (V-I) \le 1.2$, as displayed in Figure 8.
Figure 8 shows
that there is a strong single peak above the slowly varying background
in the $I$-band luminosity functions in all the regions.
We measure the peak magnitude of the red clumps
fitting the $I$-band luminosity functions
with the combination of a gaussian function (for the red clump) and
a parabolic function
(for the red giants and subgiants) as follows \citep{pac98}:

\begin{equation}
N(I_{rc}) =a + b(I-I_{rc}) + c(I-I_{rc})^2 \\
+ {{N_{RC} \over {\sigma_{RC} \sqrt{2\pi}} } 
 \exp{ \left[-{(I-I_{rc})^2} \over {2\sigma_{RC}^2} \right] }}.
\end{equation}

Figure 8 shows that the data for the $I$-band luminosity functions 
are well fit by the equations (the solid lines),
and the fitting parameters thus derived are listed in Table 4.
For the R14 and R12 regions in Figure 8,
in which only the stars in the PC chip are used because of crowding,
the luminosity functions are arbitrarily multiplied by a factor of 5
for plotting.
Remarkably, the peak magnitudes of the red clumps of all the regions
turn out to be within a very narrow range from $I_{RC} = 24.46$ to 
24.57.

Then we need to know the absolute magnitude of the red clump to 
derive the distance modulus.
Calibration of the absolute magnitude of the red clump is  a matter of
considerable debate (see \citet{gir01} and references therein). As a 
result, the calibrations of the relation between the $I$-band magnitude
and metallicity of the red clump continue to change. In fact, these 
relations should include the effects of age, but rarely do because of 
the difficulty in estimating the ages of individual field stars.
Two of the most recent empirical calibrations are

\begin{equation}
M_I^{RC}=0.19[\pm0.05][Fe/H]-0.23[\pm0.03]
\end{equation}
\noindent given by \citet{pop00} 
with the metallicity range of $-1.9\leq[Fe/H]\leq0.0$ , and

\begin{equation}
M_I^{RC}=0.13[\pm0.07][Fe/H]-0.23[\pm0.02].
\end{equation}

\noindent given by \citet{uda00} with the metallicity range of 
$-0.6\leq[Fe/H]\leq0.05$.

Assuming that the mean metallicity of the red clump in all the 
regions in our study
is similar to that of the red giants
and that the eqs.(6) and (7) are universal,
we derive the distance to M33 
using the above calibrations.
The results thus obtained are summarized in Table 4.
Table 4 lists the estimated parameters related to the red clump 
method;
the peak magnitude of the red clump ($I_{RC}$),
the dereddened peak magnitude of the red clump ($I_{0,RC})$,
the dispersion of the peak magnitude of the red clump ($\sigma_{RC}$),
the mean color of the red clump [$(V-I)_{0,RC}]$, 
the metallicity ($[Fe/H]$) (from Table 2),
the distance modulus from equation (6),
and the distance modulus from equation (7).

We have investigated the relations between these parameters in 
Figure 9 wherein (a) illustrates the mean color of the red clump stars 
as a function of [Fe/H].
This figure shows that the mean color of the red clump is almost 
constant among the regions,
with an average value, $(V-I)_{RC}=0.98 \pm 0.02$ (represented by the 
dashed line). 

Figure 9(b) displays $\sigma_{RC}$ versus [Fe/H].
$\sigma_{RC}$ is almost constant at $0.24\pm0.02$ for [Fe/H] $<-0.7$ 
dex, and
increases significantly for [Fe/H] $>-0.7$ dex.
This increase is mainly due to the crowding
problem  in the inner regions of M33 with [Fe/H] $>-0.7$ dex.

Figure 9(c) illustrates the extinction-corrected
$I$-band magnitude of the red clump versus [Fe/H].
It is clear that the $I$-band magnitude of the red clump becomes
fainter as [Fe/H] increases.
The data for all the regions are fit well by
$I_{0,RC} = 0.33[\pm0.10]$[Fe/H]$+24.67[\pm0.07]$
(leading to $M^I_{RC} = 0.33[\pm0.10]$[Fe/H]$-0.14[\pm0.07]$
when $(m-M)_{0}=24.81$ is adopted.
If using only the outer 8 regions for the fit, the relations are
$I_{0,RC} = 0.28[\pm0.14]$[Fe/H]$+24.63[\pm0.10]$
($M^I_{RC} = 0.28[\pm0.14]$[Fe/H]$ -0.18[\pm0.10]$ when
$(m-M)_0=24.81$ is adopted). 
The slope in this fit is somewhat steeper than
that of the calibration given by \citet{pop00}, 0.19,
(the dashed line).
Noting that the result on the slope is based on the assumption that the metallicity 
of the RGB in M33 is the same as that of the RC, 
determination of the metallicity of the RC is needed to confirm it.

If we normalize the fitting result based on the outer 8 regions to
the local Hipparcos red clump result of $M^I_{RC}=-0.23\pm0.02$
at $[Fe/H]=0.0$ \citep{sta98a}, 
then the relation between $M^I_{RC}$ and $[Fe/H]$ is derived 
as follows.

\begin{equation}
M_I^{RC}=0.28[\pm0.14]([Fe/H]-0.18)-0.18[\pm0.02].
\end{equation}

The distance modulus obtained using eq.(8) is 
$(m-M)_{0}=24.86\pm0.04$, which is slightly larger
than the distance moduli based on \citet{pop00}'s and 
\citet{uda00}'s calibration of $M^I_{RC}$ (see below). 

As emphasized earlier, models predict that $M^{I}_{RC}$ depends 
on metallicity and age \citep{col98, gir98, gir00, gir01}. 
The sensitivity to metal abundance is explicitly shown in Fig. 9(c) and 
quantified in the discussion above. The dependence on age is likely 
to be manifested in the dispersion around the dotted line in Fig. 9(c). 
This dispersion amounts to a root-mean-square deviation of the points 
from the fit of 0.03 mag, which is larger than the typical error in 
$I_{0,RC}$ of $\sim$0.01 mag. If taken at face value, this represents 
an age dispersion of $\sim$1.5 Gyr among the RC stars (based 
on 0.02 mag/Gyr from the models presented by \citet{sar99}).

The mean value of the distance moduli for ten regions is derived to be
$(m-M)_{0,RC} = 24.80\pm0.04$(random)$\pm0.05$(systematic) 
using eq.(6).
The errors are derived following the error budget in Table 5.
If we use the calibration by \citet{uda00},
we obtain $(m-M)_{0,RC}=24.76\pm0.04$(random)$\pm0.05$(systematic).
These values are in excellent agreement with those from the TRGB.

\section{DISCUSSION}

\subsection{Comparison with Other Studies}

To date the distance to M33 has been studied using a number of standard 
candles: Cepheid variables \citep{san83a, san83b, chr87, mou87, mad91, 
fre91,fre01,lee01},
horizontal branch stars in globular clusters \citep{sar00},
red supergiant long-period variables (SLPVs)\citep{pie00},
the luminosity function of the planetary nebulae (PNLF)\citep{mag00} 
and
the TRGB \citep{mou86,lee93,sal98}, as summarized in Table 6
(see also \citet{van91},\citet{van99},\citet{van00}).
These distance moduli range from as low as 24.41 to a 
high of 24.85. 
Our values of
24.80, 24.81 (from the RC with Popowski (2000) calibration and the TRGB)
and  24.76 (from the RC with Udalski (2000) calibration) 
are at the high end of the published range of distances.

Among the previous distance estimates, 
\citet{lee01} determined the distance to M33 using the
single phase $I$-band photometry of 21 Cepheids with $\log P >0.8$ 
based on the same data as used in this paper.
\citet{lee01} obtained $(m-M)_0=24.52\pm0.13$ for
for an adopted total 
reddening of M33, $E(B-V)=0.20\pm0.04$ ($E(V-I)=0.27\pm0.05$) given by
\citet{fre01}, 
the reddening to the LMC, $E(B-V)=0.10$,
and the distance to the LMC, $(m-M)_0=18.50$.
This value is $\sim 0.3$ mag smaller than those derived using the TRGB and RC
in this study. This difference is considered partially due to the uncertainty
in the estimates of the total reddening for Cepheids in M33.
Note that  \citet{fre91} derived the total reddening of M33 Cepheids from
$BVRI$ photometry to be $E(B-V)=0.10\pm0.09$, while \citet{fre01} revised this value
to $E(B-V)=0.20\pm0.04$ using the different period-luminosity relations for
$V$ and $I$ with the same data. 
Better estimates of the reddening of M33 Cepheids are needed to 
investigate further this problem.

\subsection{Magnitude difference between the TRGB and the RC}

In the previous section we have examined the dependence on metallicity
of the $I$-band magnitude of the red clump and the magnitude of the 
TRGB.
Here we investigate the dependence on metallicity of both together 
using
the difference of the $I$-band magnitude between the RC and the TRGB,
$\Delta I$(RC--TRGB) $ = I_{RC} - I_{TRGB}$.
$\Delta I$(RC--TRGB) can be measured directly from the photometry
and has the added advantage of being extinction-free \citep{ber00}.
Figure 10 displays $\Delta I $(RC--TRGB) versus [Fe/H] for the ten 
regions in M33.
It is seen clearly that there is a positive correlation between
$\Delta I $(RC--TRGB) and [Fe/H].
The data for the outer 8 regions in M33 are fit well by
$\Delta I $(RC--TRGB)$=0.45[\pm0.18]$[Fe/H]$+3.95[\pm0.14]$.
If we use the data for all regions including the inner two regions, 
we derive
$\Delta I $(RC--TRGB)$=0.56[\pm0.15]$[Fe/H]$+4.04[\pm0.11]$.
The error for the slope is rather large, because the range of [Fe/H] 
used for this fit is small.
Since the TRGB magnitudes and foreground extinctions for all the 
regions are almost constant, $M_{I}^{TRGB}=-4.0$ and $A_I=0.08$
(as given in Table 2),
the slope in this fit represents basically the dependence
of the RC magnitude on [Fe/H].
The slope derived from the data of M33 is rather steeper than
the slope given
by \citet{pop00} which is based on the galactic RC stars
(shown by the dashed line in Figure 10).
For a better determination of the dependence of $\Delta I $(RC--TRGB) 
on [Fe/H],
a large range of [Fe/H] is required.

We have compared $\Delta I $(RC--TRGB) for M33 with those for other 
nearby galaxies compiled by \citet{ber00} in Figure 11.
For M33 the mean difference $<\Delta I>=3.62\pm0.05$ and mean 
metallicity
$<[Fe/H]>=-0.75\pm0.07$ of the outer eight regions
 (excluding R14 and R12) are used from this study.
Figure 11 shows that the data for M33 is consistent with those for
other galaxies, following the relation plotted in Figure 10.

\section{SUMMARY}

 We present $VI$ photometry of field stars in ten regions located
 at $R=2.6$ to 17.8 arcmin from the center of M33 based on 
$HST/WFPC2$ images.
From this photometry we have determined the distance to M33
using the tip of the red giant branch (TRGB) and the red clump (RC).
Main results obtained in this study are summarized as follows.

\begin{enumerate}
\item Mean metallicities of the RGB in ten regions
range from [Fe/H]=--0.9 to --0.6.
We find a clear negative radial gradient
of the metallicity of the RGB, which has a smaller slope
and much smaller scatter than that derived from HII regions in M33.

\item $I$-band magnitudes of the TRGB in ten regions are almost 
constant with a very narrow range: $I=20.82$ to 20.92.
This result confirms that the $I$-band magnitudes of the TRGB is 
insensitive to age or metallicity for old stars with [Fe/H]$<-0.7$ \citep{lee93}.

\item The distance to M33 based on the TRGB of ten regions is derived 
to be
$(m-M)_{0,TRGB}=24.81\pm0.04$(random)$^{+0.15}_{-0.11}$(systematic)
(corresponding to a distance of $916\pm17$(random) kpc).

\item Mean colors of the RC in ten regions are almost constant: 
$(V-I)_0=0.89$ to 0.97, and show little correlation with [Fe/H].

\item $I$-band magnitudes of the RC in ten regions are almost constant with
a very narrow range: $I=24.46$ to 24.57, but they show a correlation 
with [Fe/H] with a slope similar to that given by \citet{pop00}.

\item Assuming the metallicity of the RC stars is the same as that of the
RGB stars, the distance to M33 based on the RC of ten regions is derived to be
$(m-M)_{0,RC} = 24.80\pm0.04$(random)$\pm0.05$(systematic)
(corresponding to a distance of $912\pm17$(random) kpc), which is
in excellent agreement with the TRGB distance obtained in this study.

\end{enumerate}

\acknowledgements
M.G.L. is in part supported by the MOST/KISTEP International 
Collaboration Research Program (1-99-009). 
M.G.L. is grateful to the Astronomy Group
at the University of Concepcion for the warm hospitality during his stay for this work.
A.S. has benefited from financial support from NSF CAREER grant No. AST-0094048.
D.G. acknowledges financial support for this project received from CONICYT 
through Fondecyt grant 8000002.

\clearpage

\begin{deluxetable}{rllrrc}
\tablecaption{A LIST OF THE REGIONS IN M33 USED FOR ANALYSIS 
\label{tbl-1}} \tablewidth{0pt}
\tablecolumns{11} \tablehead{ \colhead{Region}&
\colhead{R.A.(2000) \tablenotemark{a}}&
\colhead{Dec.(2000) \tablenotemark{b}} &
\colhead{R\tablenotemark{c}} &
\colhead{$R_{dp}$\tablenotemark{d}}&
\colhead{$E(V-I)$\tablenotemark{e}}\\}
\startdata
 R14 & 1 34  \phantom{0}0.80 & +30 41 \phantom{0}3.21 &  2.55 &  \phantom{0}3.26 &  0.056 \\
 R12 & 1 34  \phantom{0}5.62 & +30 38 40.01 &  3.28 &  \phantom{0}5.84 &  0.057 \\
 U77 & 1 33 26.70 & +30 41 17.72 &  5.50 &  9.80 &  0.056 \\
 U49 & 1 33 47.30 & +30 47 54.56 &  8.34 & 10.21 &  0.055 \\
  M9 & 1 34 32.40 & +30 38 24.50 &  8.98 & 15.67 &  0.061 \\
 H10 & 1 33 32.68 & +30 48 58.39 & 10.15 & 14.82 &  0.055 \\
 H38 & 1 33 49.76 & +30 28 57.86 & 10.65 & 12.13 &  0.061 \\
U137 & 1 33 13.33 & +30 28 51.51 & 13.47 & 14.32 &  0.060 \\
 C20 & 1 34 42.54 & +30 52 41.41 & 17.12 & 18.70 &  0.062 \\
 C38 & 1 33 28.31 & +30 22 28.22 & 17.83 & 18.12 &  0.063 \\
\enddata
\tablenotetext{a}{Right ascension in units of hours, minutes, and 
seconds.}
\tablenotetext{b}{Declination in units of degrees, arcminutes, and 
arcseconds.}
\tablenotetext{c}{Radial distance in arcminutes from the center of 
M33.}
\tablenotetext{d}{Deprojected radial distance in arcminutes from the 
center of M33.}
\tablenotetext{e}{Foreground reddening from COBE/IRAS maps of 
\citet{sch98}.}
\end{deluxetable}

\begin{deluxetable}{rccccccc}
\tablecaption{
ESTIMATED PARAMETERS FOR THE TIP OF THE RED GIANT BRANCH
METHOD\label{tbl-2}} \tablewidth{0pt}
\tablecolumns{11}\tablehead{\colhead{Region}&
\colhead{$I_{TRGB}$}&
\colhead{$I_{0,TRGB}$}&
\colhead{$(V-I)_{0,TRGB}$}&
\colhead{$(V-I)_{0,-3.5}$}&
\colhead{[Fe/H]}&
\colhead{$M_{I,TRGB}$}&
\colhead{$(m-M)_{0,TRGB}$}\\}
\startdata
  R14 & 20.82 $\pm$  0.05 & 20.74 &  2.50 $\pm$  0.06 &  1.89 $\pm$  0.01 & --0.61 $\pm$  0.09 & --3.97 & 24.71 \\
  R12 & 20.87 $\pm$  0.05 & 20.79 &  2.35 $\pm$  0.06 &  1.81 $\pm$  0.01 & --0.65 $\pm$  0.09 & --4.00 & 24.79 \\
  U77 & 20.87 $\pm$  0.05 & 20.79 &  2.35 $\pm$  0.06 &  1.76 $\pm$  0.01 & --0.68 $\pm$  0.09 & --3.99 & 24.78 \\
  U49 & 20.87 $\pm$  0.05 & 20.80 &  2.35 $\pm$  0.05 &  1.78 $\pm$  0.01 & --0.67 $\pm$  0.09 & --3.99 & 24.79 \\
   M9 & 20.92 $\pm$  0.05 & 20.84 &  2.05 $\pm$  0.05 &  1.68 $\pm$  0.02 & --0.79 $\pm$  0.10 & --4.04 & 24.88 \\
  H10 & 20.92 $\pm$  0.05 & 20.84 &  2.30 $\pm$  0.05 &  1.73 $\pm$  0.02 & --0.71 $\pm$  0.09 & --4.00 & 24.84 \\
  H38 & 20.82 $\pm$  0.05 & 20.74 &  2.15 $\pm$  0.05 &  1.73 $\pm$  0.01 & --0.72 $\pm$  0.09 & --4.03 & 24.77 \\
 U137 & 20.87 $\pm$  0.05 & 20.79 &  2.25 $\pm$  0.05 &  1.71 $\pm$  0.02 & --0.74 $\pm$  0.09 & --4.00 & 24.79 \\
  C20 & 20.87 $\pm$  0.05 & 20.78 &  2.05 $\pm$  0.05 &  1.66 $\pm$  0.02 & --0.82 $\pm$  0.09 & --4.04 & 24.82 \\
  C38 & 20.92 $\pm$  0.05 & 20.83 &  2.00 $\pm$  0.05 &  1.63 $\pm$  0.01 & --0.86 $\pm$  0.09 & --4.04 & 24.88 \\
\enddata
\end{deluxetable}

\begin{deluxetable}{llr}
\tablecaption{ERROR BUDGET FOR THE TRGB METHOD \label{tbl-3}} 
\tablewidth{0pt}
\tablecolumns{3} \tablehead{ \colhead{}&
\colhead{Error}&
\colhead{Estimation (mag)}\\}
\startdata
1. &Random Error& \\
   &A. Reddening & 0.03 \\
   &B. Photometry & 0.008 \\
   &C. Tip detction & 0.05 \\
   &D. Color spread in the RGB population & 0.02 \\
   &E. HST photometric calibration & 0.12 \\
   &F. Total & 0.135 \\
   &G. Total for 10 regions & 0.04\tablenotemark{a} \\
   &H. Total for 6 regions & 0.06\tablenotemark{a} \\
\hline
2. &Systematic Error & \\
   &A. RR Lyrae distance scale & 0.11 \\
   &B. Undersampling in the Galactic globular cluster calibration\tablenotemark{b} &
0.1 \\
   &C. Total & $^{+0.15}_{-0.11}$ \\
\hline
\enddata
\tablenotetext{a}{$G~or~H=F/\sqrt{number~of~regions}$ 
which is included the statistical errors.}
\tablenotetext{b}{This error is propagated in only one direction which 
underestimates the TRGB brightness.}
\end{deluxetable}

\begin{deluxetable}{rccccccc}
\tablecaption{ESTIMATED PARAMETERS FOR THE RED CLUMP 
METHOD\label{tbl-4}} \tablewidth{0pt}
\tablecolumns{11}\tablehead{\colhead{Region}&
\colhead{$I_{RC}$}&
\colhead{$I_{0,RC}$}&
\colhead{$\sigma_{RC}$}&
\colhead{$(V-I)_{0,RC}$\tablenotemark{a}}&
\colhead{$[Fe/H]$\tablenotemark{b}}&
\colhead{$(m-M)_{0,RC}$\tablenotemark{c}}&
\colhead{$(m-M)_{0,RC}$\tablenotemark{d}}\\}
\startdata
  R14 & 24.57 $\pm$  0.02 & 24.49 &  0.33 $\pm$  0.09 &  0.93 $\pm$ 0.006 & --0.61 $\pm$  0.09 & 24.84  & 24.80 \\
  R12 & 24.52 $\pm$  0.02 & 24.44 &  0.41 $\pm$  0.19 &  0.89 $\pm$ 0.006 & --0.65 $\pm$  0.09 & 24.80  & 24.76 \\
  U77 & 24.49 $\pm$  0.01 & 24.41 &  0.38 $\pm$  0.07 &  0.92 $\pm$ 0.004 & --0.68 $\pm$  0.09 & 24.77  & 24.73 \\
  U49 & 24.52 $\pm$  0.01 & 24.44 &  0.34 $\pm$  0.06 &  0.93 $\pm$ 0.004 & --0.67 $\pm$  0.09 & 24.79  & 24.75 \\
   M9 & 24.49 $\pm$  0.01 & 24.40 &  0.24 $\pm$  0.01 &  0.94 $\pm$ 0.004 & --0.79 $\pm$  0.09 & 24.78  & 24.74 \\
  H10 & 24.56 $\pm$  0.01 & 24.48 &  0.26 $\pm$  0.02 &  0.97 $\pm$ 0.004 & --0.71 $\pm$  0.09 & 24.84  & 24.80 \\
  H38 & 24.53 $\pm$  0.01 & 24.44 &  0.25 $\pm$  0.02 &  0.89 $\pm$ 0.004 & --0.72 $\pm$  0.09 & 24.81  & 24.76 \\
 U137 & 24.53 $\pm$  0.01 & 24.44 &  0.25 $\pm$  0.01 &  0.93 $\pm$ 0.004 & --0.74 $\pm$  0.09 & 24.81  & 24.76 \\
  C20 & 24.46 $\pm$  0.01 & 24.37 &  0.21 $\pm$  0.01 &  0.92 $\pm$ 0.004 & --0.82 $\pm$  0.09 & 24.76  & 24.71 \\
  C38 & 24.49 $\pm$  0.01 & 24.40 &  0.21 $\pm$  0.01 &  0.92 $\pm$ 0.003 & --0.86 $\pm$  0.09 & 24.80  & 24.75 \\
\enddata
\tablenotetext{a}{Values from the mean loci of RC with $I_{0,RC}$.}
\tablenotetext{b}{[Fe/H] from the TRGB method. See Table 2.}
\tablenotetext{c}{Distance modulus based on \citet{pop00}'s 
calibration.}
\tablenotetext{d}{Distance modulus based on \citet{uda00}'s 
calibration.}
\end{deluxetable}

\begin{deluxetable}{llr}
\tablecaption{ERROR BUDGET FOR THE RED CLUMP METHOD \label{tbl-5}} 
\tablewidth{0pt}
\tablecolumns{3} \tablehead{ \colhead{}&
\colhead{Error}&
\colhead{Estimation (mag)}\\}
\startdata
1. &Random error  &\\
   &A. Reddening & 0.03 \\
   &B. Photometry & 0.07 \\
   &C. Mean $I_{RC}$ detection (fitting error) & 0.01 \\
   &D. HST photometric calibration & 0.12 \\
   &E. Total & 0.14 \\
   &F. Total for 10 regions & 0.04\tablenotemark{a} \\
\hline
2. &Systematic error\tablenotemark &\\
   &A1. calibration error in Powposki (2000) & 0.05 \\
   &A2. calibration error in Udalski (2000) & 0.05 \\
\hline
\enddata
\tablenotetext{a}{$F~=~E/\sqrt{number~of~regions}$ which includs the
statistical error.}
\end{deluxetable}

\begin{deluxetable}{rcccl}
\tablecaption{A LIST OF M33 DISTANCE ESTIMATES \label{tbl-6}} 
\tablewidth{0pt}
\tablecolumns{11}\tablehead{\colhead{Method}&
\colhead{$(m-M)_{0}$}&
\colhead{$(m-M)_{0}$\tablenotemark{a}}&
\colhead{$E(B-V)$}&
\colhead{Reference}\\}
\startdata
Cepheids $m_{pg}$ & $24.05\pm0.18$                  &$24.32$ & 
  0.12                  & Christian \& Schommer (1987) \\
Cepheids I        & $24.82\pm0.15$                  &$24.93$ & 
  0.12                  & Mould (1987) \\
Cepheids BVRI     & $24.64\pm0.09$                  &$24.75$ & 
  0.10                  & Freedman, Wilson, \& Madore (1991)\\
Cepheids VI     & $24.56\pm0.10$                  &$24.87$ & 
  0.20                  & Freedman et al. (2001)\\
Cepheids VI       & $24.52\pm0.15$                  &$24.83$ & 
  0.20                  & Lee et al. (2001) \\
TRGB     & $24.70\pm0.10$                  &$24.82$ & 
  0.10                  & \citet{mou86},\citet{lee93}\\
SLPV     & $24.85\pm0.13$                  &$24.97$ & 
  0.10                  & Pierce, Jurcevic, \& Crabtree(2000)\\
PNLF     & $24.62\pm0.25$                  &$24.74$ & 
  0.10                  & Magrini et al. (2000)\\
HB       & $24.84\pm0.16$                  & ---    & 
  0.04\tablenotemark{b} & Sarajedini et al. (2000)\\
TRGB     & $24.81\pm0.13$                  & ---    & 
  0.04\tablenotemark{b} & This study\\
RC       & $24.80\pm0.14$\tablenotemark{c} & ---    & 
  0.04\tablenotemark{b} & This study\\
RC       & $24.76\pm0.14$\tablenotemark{d} & ---    & 
  0.04\tablenotemark{b} & This study\\
\enddata
\tablenotetext{a}{Distance modulus for $E(B-V)=0.04$.}
\tablenotetext{b}{The foreground reddening value of \citet{sch98}.}
\tablenotetext{c}{Based on \citet{pop00}'s calibration.}
\tablenotetext{d}{Based on \citet{uda00}'s calibration.}
\end{deluxetable}

\clearpage
\begin{figure}
\epsscale{0.5}
\plotone{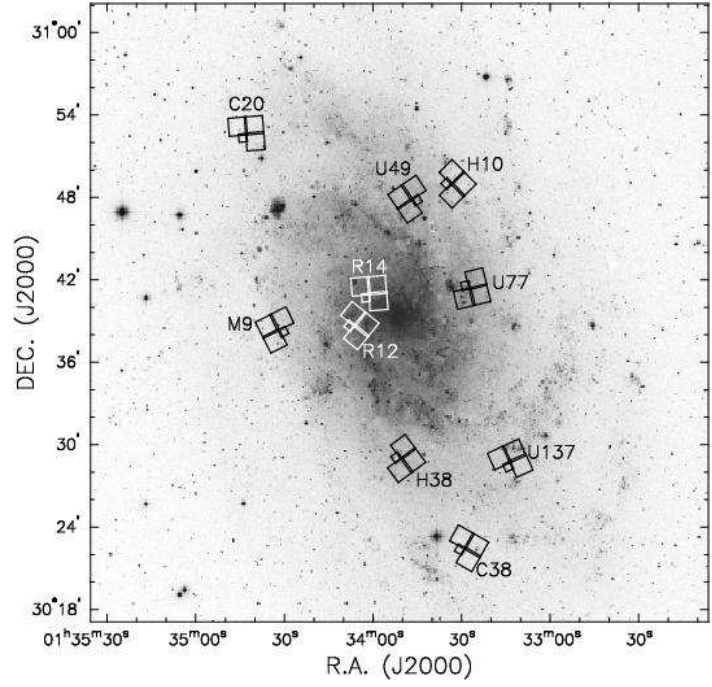}
\figcaption[fig01.ps]{
A finding chart of M33 showing the positions of the ten regions 
observed with $HST/WFPC2$
(squares). The grayscale map is from the digitized Palomar Sky Survey.
\label{fig1}}
\end{figure}

\begin{figure}
\epsscale{0.6}
\plotone{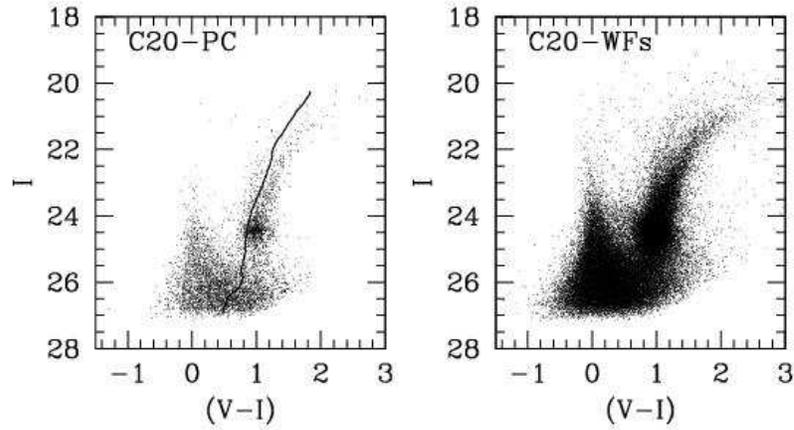}
\figcaption[fig02.ps]{
Color-magnitude diagrams of the field stars in the PC chip (left panel) and
in the WF2, WF3, and WF4 chips (right panel) of the C20-region.
The solid line in the left panel represents the mean locus of the 
globular cluster C20.
Note that the mean color of the RGB of the field stars is redder than 
that of the globular cluster.
A compact red clump is found to be at $I \approx 24.4$ and
at $(V-I) \approx 0.9$, and the TRGB is seen at $I \approx 21.0$.
\label{fig2}}
\end{figure}

\begin{figure}
\epsscale{1}
\plotone{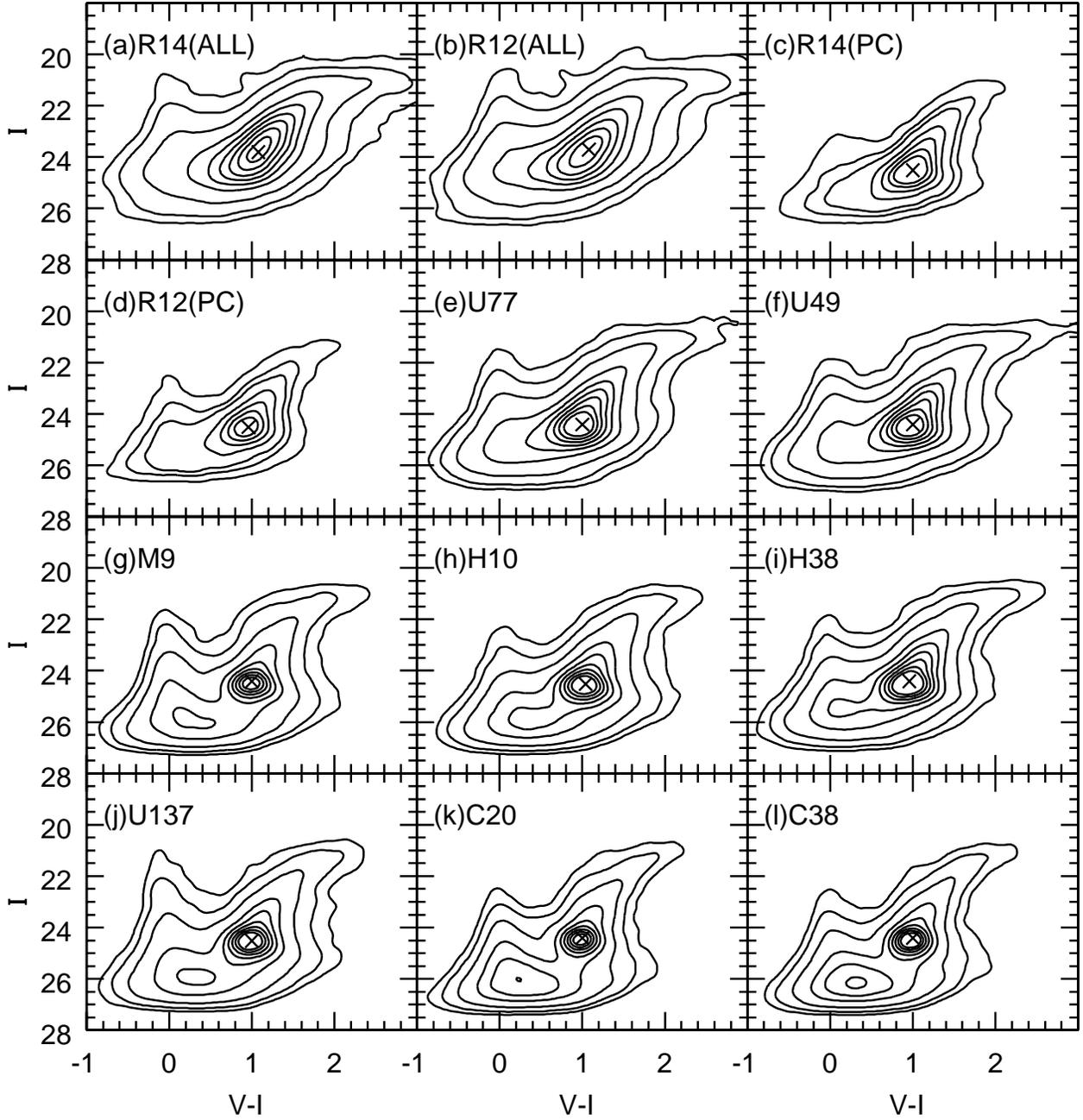}
\figcaption[fig03.ps]{
Color-magnitude diagrams of the observed regions in M33 
displayed in the density contour map.
The panels (c) and (d), for R14 and R12 regions, are only for the stars
in the PC chip. In other panels the CMDs of the field stars in all 
four chips (PC, WF2, WF3, and WF4) are displayed.
Crosses represent the peaks of the density contour map, which 
correspond to the mean position of the red clumps. 
Contour levels are at 1, 3, 10, 30, 50, 70, 90, 110, and 130 stars/grid, respectively.
For (c) and (d) contour levels are at 1, 3, 5, 10, 15, 20, and 25 
stars/grid, respectively.
\label{fig3}}
\end{figure}

\begin{figure}
\epsscale{1}
\plotone{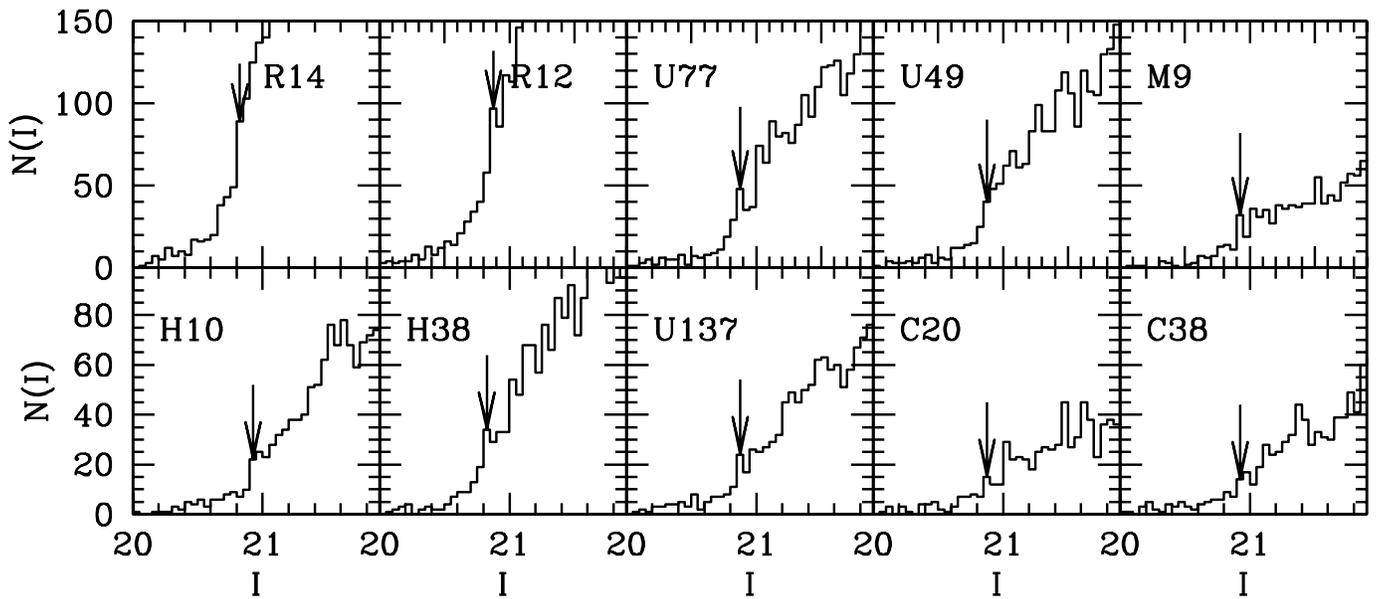}
\figcaption[fig04.ps]{
$I$-band luminosity functions of bright red stars in M33.
The $I$ magnitudes of the tip of the red giant branch are marked by 
arrows.
\label{fig4}}
\end{figure}

\begin{figure}
\epsscale{0.5}
\plotone{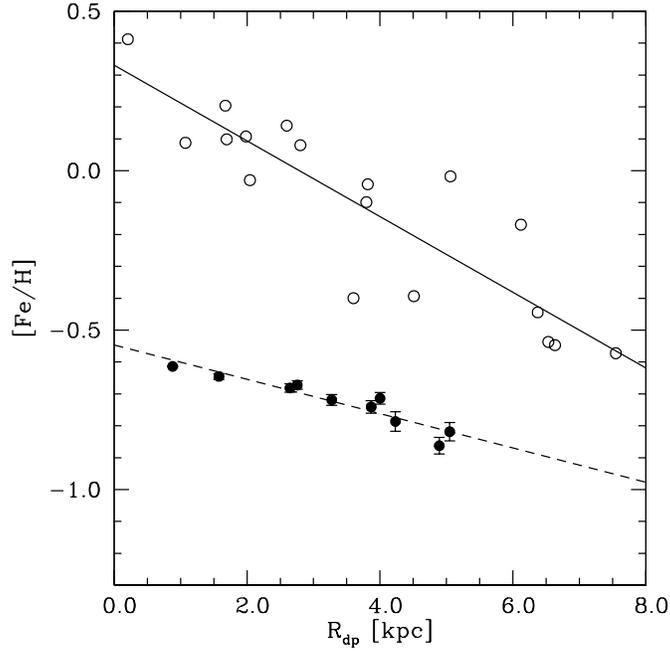}
\figcaption[fig05.ps]{
Mean metallicity [Fe/H] of the RGB in each region 
as a function of the deprojected radial distance derived in this 
study (filled circles).
The dashed line represents the mean relation between $R_{dp}$ and
[Fe/H] derived from the RGB.
The mean metallicity decreases as the radial distance increases.
For comparison, the solid line represents the mean relation
between $R_{dp}$ and [Fe/H]
derived for the HII regions in M33 from various other studies (open 
circles).
\label{fig5}}
\end{figure}

\begin{figure}
\epsscale{0.5}
\plotone{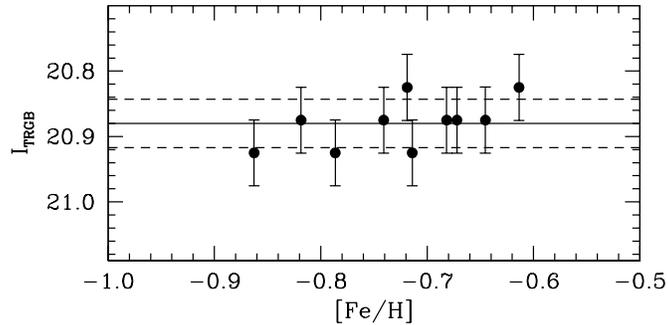}
\figcaption[mskim.fig06.ps]{
The $I$-band magnitude of the TRGB, $I_{TRGB}$,
as a function of metallicity. 
The solid line represents the average value of $I_{TRGB}$, and
the dashed lines represent standard deviations of $\pm1\sigma$.  
\label{fig6}}
\end{figure}

\begin{figure}
\epsscale{1}
\plotone{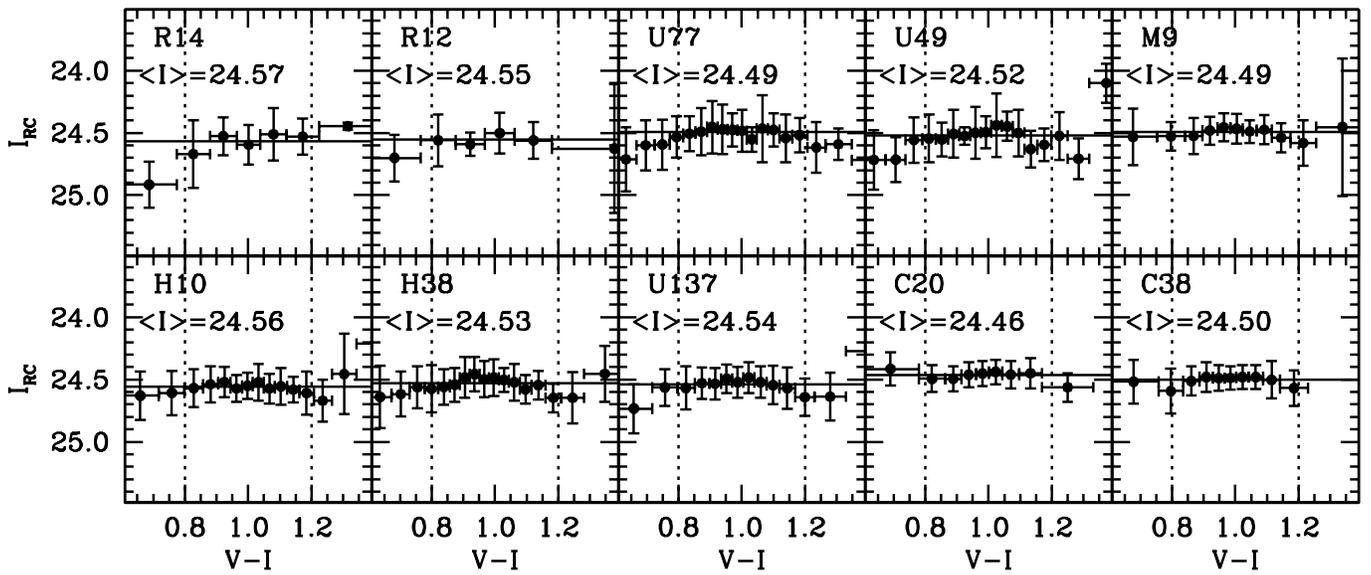}
\figcaption[fig07.ps]{
The mean magnitude of the red clump $I_{RC}$ versus
$(V-I)$ color. The horizontal bars on filled circles represent the sizes of
color bins, not error bars, while the vertical bars represent fitting 
errors.  
Two thousand stars are included in each color bin.
Note that $I_{RC}$ varies little
in the color range of $0.8\le(V-I)\le1.2$ (between the two vertical 
dotted lines).
The horizontal solid line represents the mean of $I_{RC}$ for all 
color bins.
\label{fig7}}
\end{figure}

\begin{figure}
\epsscale{1}
\plotone{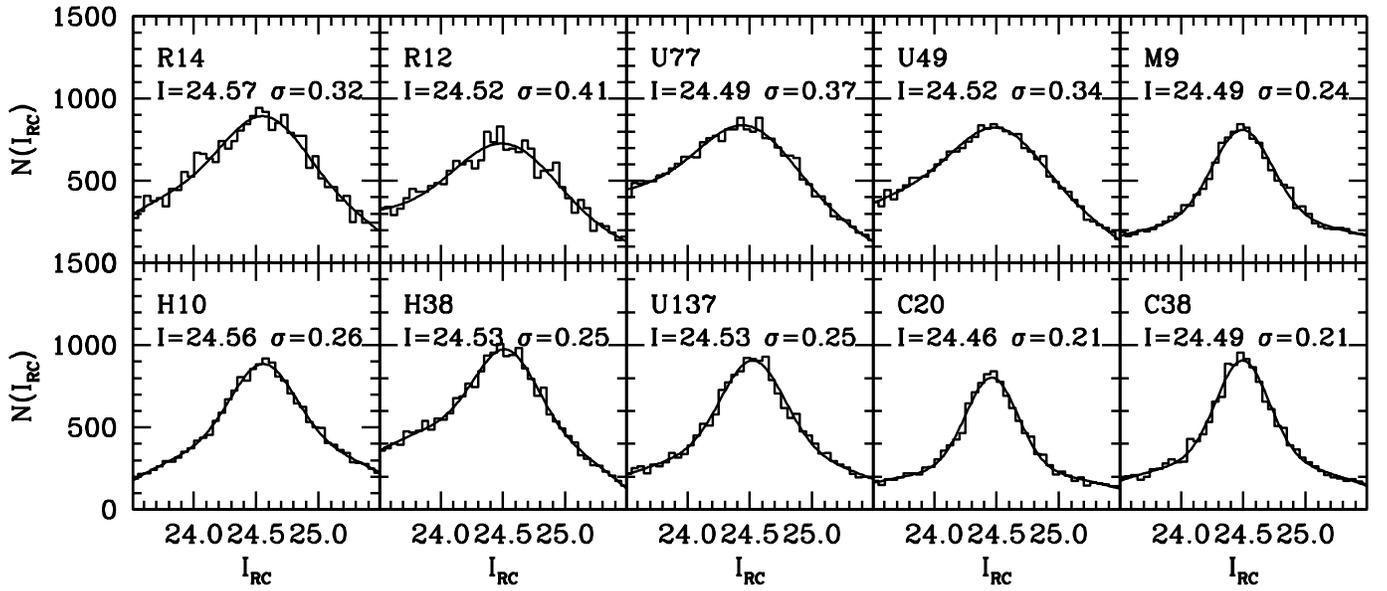}
\figcaption[fig08.ps]{
$I$-band luminosity functions of red clump stars in the color range
$0.8 < (V-I) < 1.2$ and in the magnitude
range $23.5 < I < 25.5$. The solid line is a
fitting line with equation (5) as described in the text.
The mean magnitude of the red clump $I$ and the dispersion of $I$,
$\sigma$, are also shown.
The luminosity functions of R14 and R12 regions, in which the stars in only 
PC chips are used,
are multiplied by a factor of 5 arbitrarily.
\label{fig8}}
\end{figure}

\begin{figure}
\epsscale{0.5}
\plotone{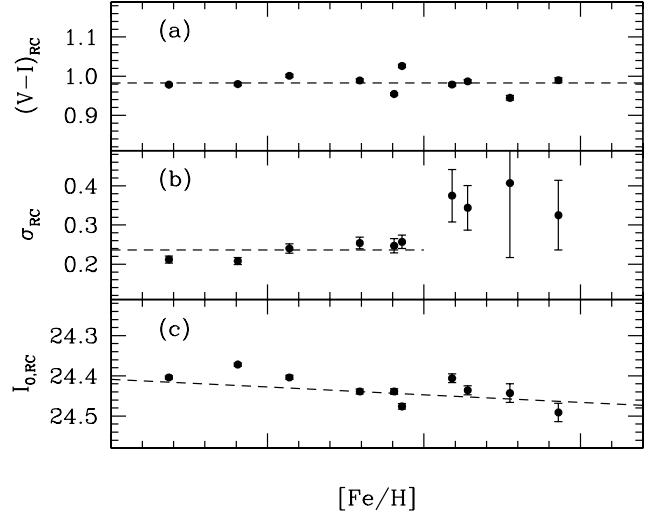}
\figcaption[fig09.ps]{
Several parameters for the red clump method as a function of 
metallicity [Fe/H].
(a) The mean color of the red clump $(V-I)_{RC}$. 
The dashed line represents a mean value of the ten regions. 
(b) The dispersion of the mean magnitude of red clump. 
The dashed line represents a mean value of the outer six regions with 
lower [Fe/H].
(c) The extinction corrected $I$-band magnitudes of the red clump. 
The dashed line shows the \citet{pop00}'s calibration 
for a distance modulus of $(m-M)_0 =24.81$.
\label{fig9}}
\end{figure}

\begin{figure}
\epsscale{0.5}
\plotone{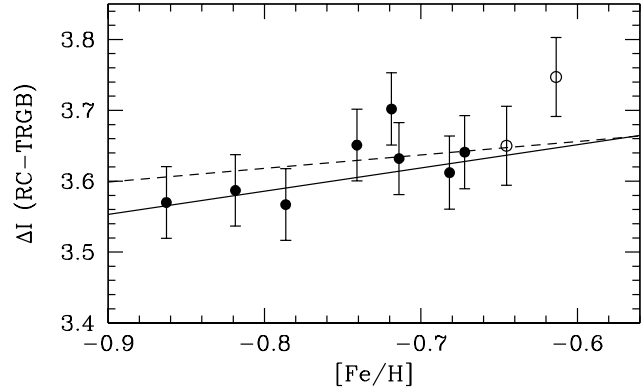}
\figcaption[fig10.ps]{
The difference in $I$-band magnitude between the RC and the TRGB, 
$\Delta I$(RC--TRGB),
 versus mean metallicity of the red giants in M33.
The dashed line represents the expected difference,
when the TRGB magnitude $M_{I}^{TRGB}=-4.0$ and \citet{pop00}'s 
calibration 
are assumed. 
The solid line represents a fitting line to the data when R14 and 
R12 regions 
(open circles) are excluded. 
\label{fig10}}
\end{figure}

\begin{figure}
\epsscale{0.5}
\plotone{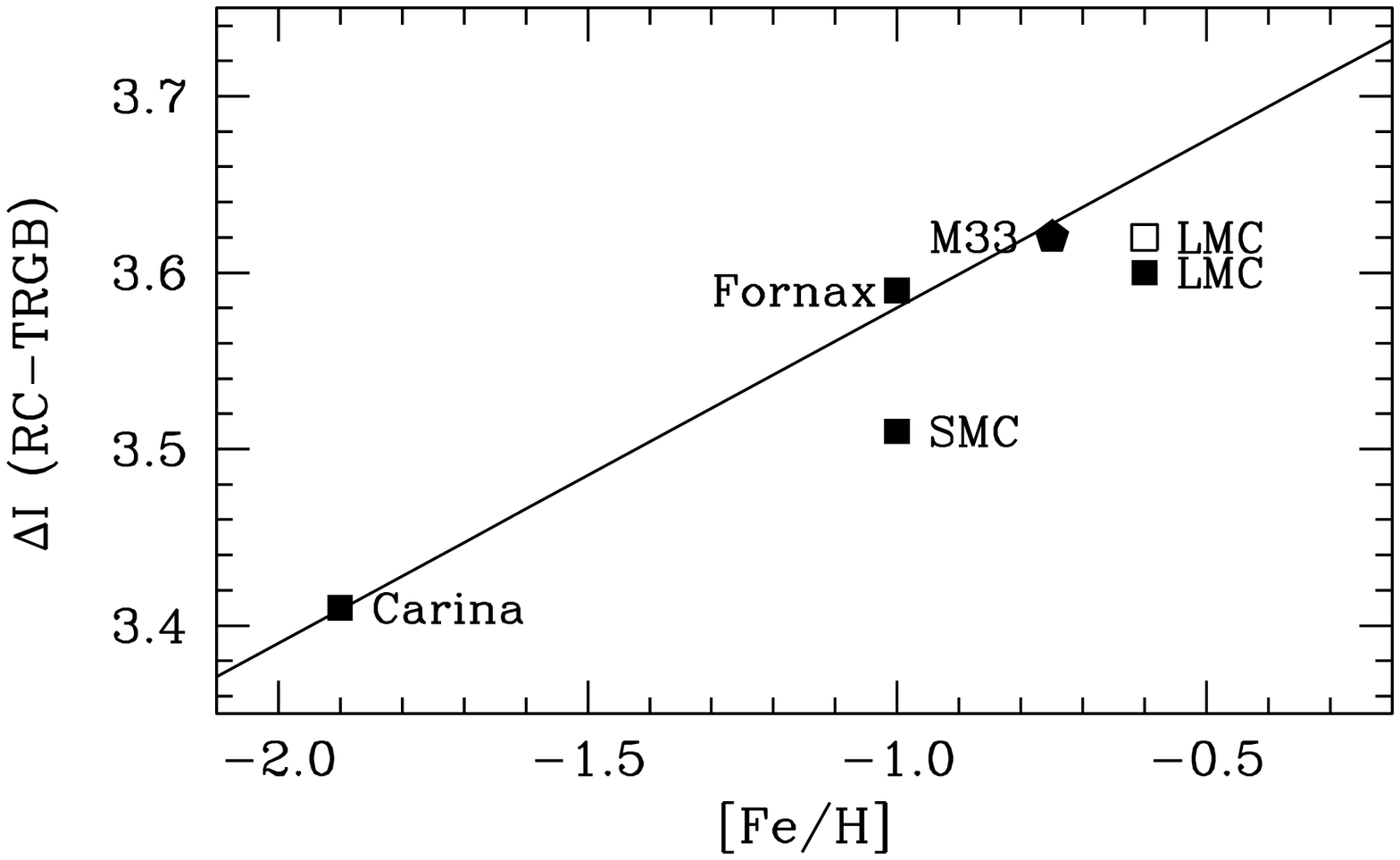}
\figcaption[fig11.ps]{
The difference in $I$ magnitude between the RC and the TRGB stars
$\Delta I $(RC--TRGB) versus metallicity for M33 and other nearby 
galaxies.
A filled pentagon represents our mean value of the eight regions in 
M33.
The LMC data are from \citet{zar97} (open square) and 
\citet{sak00} (filled square).
The solid line represents the expected difference
when  the TRGB magnitude $M_{I}^{TRGB}=-4.0$ and \citet{pop00}'s 
calibration are assumed.
\label{fig11}}
\end{figure}

\clearpage
\end{document}